\documentclass[amsmath,amssymb,aps,pre,showpacs,twocolumn,superscriptaddress]{revtex4-2}
\usepackage[dvipsnames,table,xcdraw]{xcolor}
\usepackage[utf8]{inputenc}
\usepackage{hyperref}
\usepackage{wrapfig}
\usepackage{bm}
\usepackage[capitalise]{cleveref}
\usepackage{float}
\usepackage{graphicx,epstopdf,subfigure}
\usepackage{hyperref}
\usepackage{amsmath}  
\usepackage{amssymb}
\usepackage{gensymb}
\usepackage{enumitem}
\usepackage{indentfirst}
\usepackage{blkarray, bigstrut}
\usepackage{rotating}
\newcommand{\add}[1]{\textcolor{black} {{#1}}}

\begin{abstract}
\add{Full synchronization of dynamical elements coupled via hypergraphs can be analyzed with the hypergraph projection onto dyadic matrices, but this is not sufficient for analyzing cluster synchronization. Here we develop the necessary formalism.} We introduce the notion of edge clusters and show how node and edge partitions allow us to verify admissible states and simplify their linear stability calculations. This provides a principled way to track dynamics on hypergraphs, and the projected Laplacian matrices based on each edge cluster are essential to \add{linear stability analysis and its dimensionality reduction}. This work goes beyond full synchronization and beyond dyadic interactions. 

\end{abstract}
\begin{document}
\title{Cluster synchronization on hypergraphs}
\author{Anastasiya Salova}
\email[]{avsalova@ucdavis.edu}
\affiliation{Department of Physics and Astronomy, University of California, Davis, CA 95616, USA}
\affiliation{Complexity Sciences Center, University of California, Davis, CA 95616, USA}
\author{Raissa M. D'Souza}
\affiliation{Complexity Sciences Center, University of California, Davis, CA 95616, USA}
\affiliation{
	Department of Computer Science and
	Department of Mechanical and Aerospace Engineering, University of California, Davis, CA 95616, USA}
\affiliation{Santa Fe Institute, Santa Fe, NM 87501, USA}

\maketitle



\add{\textit{Introduction:}}
Patterns of synchronization in complex interdependent dynamical systems, from full synchronization to cluster synchronization where different groups of oscillators follow distinct synchronized trajectories,  can be essential to their function. Such systems are often modeled by networks of agents with dyadic interactions \cite{newman2018networks}. \add{Cluster synchronization on dyadic networks can manifest 
intriguing behaviors such as remote synchronization and chimera states \cite{cho2017stable,gambuzza2013analysis}
and its stability analysis  is well established \cite{pecora2014cluster}.}
However, dyadic interactions may not be sufficient.
Higher order interactions are required to describe certain 
chemical \cite{jost2019hypergraph}, biological \cite{klamt2009hypergraphs}, and coauthorship interactions \cite{han2009understanding,lung2018hypergraph}, and processes such as consensus dynamics \cite{neuhauser2020multibody,neuhauser2021consensus}, making it necessary to go beyond pairwise analysis \cite{battiston2020networks,bick2021higher,torres2020and}. A simplicial complex or a hypergraph can be used to encode the structure of these higher order interactions which can 
support diverse types of dynamics. For instance,
different 
higher-order generalizations of dyadic phase oscillator models have recently led to discovery of new behaviors 
\cite{ashwin2016hopf,bick2016chaos,skardal2019abrupt,bick2020multi,skardal2020memory,millan2020explosive}.

\add{Stability analysis of dynamics on hypergraphs 
is challenging because interactions of all orders contribute. 
Yet, full synchronization and its stability on hypergraphs has been analyzed in diverse settings, 
generalizing the master stability formalism from dyadic to higher order interactions \cite{gambuzza2020master, de2021phase, lucas2020multi,PhysRevE.101.062313,krawiecki2014chaotic,de2021phase,PhysRevE.101.062313,gambuzza2020master,carletti2020dynamical}. 
Non-intertwined cluster synchronization \cite{zhang2020unified} and cluster synchronization on chemical hypergraphs \cite{bohle2021coupled} have also been analyzed. 
Often, as with full synchronization, the projection of the hypergraph onto a set of dyadic interaction matrices, one for each order, is sufficient to simplify the stability analysis, but this is not the case 
for more intricate dynamics on hypergraphs like cluster synchronization \cite{salova2021analyzing}.   
%
Synchronization patterns can arise from symmetries, for which general methods to simplify stability analysis in systems with dyadic and non-dyadic interactions 
exist \cite{golubitsky2003symmetry,golubitsky2016rigid,pecora2014cluster}.
However, patterns of cluster synchronization can also arise from more general properties of the hypergraph structure.
}

\add{In this Letter, we introduce a framework to analyze the dynamics and stability of cluster synchronization on general hypergraphs with Laplacian-like coupling
arising from symmetries and beyond.}
This provides a systematic way to track the effective system dynamics and to identify a set of matrices that must be simultaneously block diagonalized to simplify stability calculations. We use external equitable partitions of the incidence matrix to determine admissible patterns of 
cluster synchronization \cite{belykh2008cluster,sorrentino2016complete,siddique2018symmetry,schaub2016graph}.  The results on simultaneously block diagonalization \cite{maehara2011algorithm} generalize the work of Refs.\cite{irving2012synchronization, zhang2020symmetry,pecora2014cluster} from a dyadic to a general hypergraph setting, and beyond non-intertwined clusters. 
\add{To present the formulation most cleanly and relate to existing literature \cite{lucas2020multi,gambuzza2020master,de2021phase}, we consider generalized Laplacian coupling between elements. In a companion manuscript \cite{salova2021analyzing} we consider more general undirected coupling and moreover show how the hypergraph projection onto a dyadic network fails to capture the full information 
necessary for analyzing cluster synchronization.}
We release accompanying code that can be used for admissibility and stability calculations \cite{salova2021code}. 




\add{Patterns of synchronization 
are well-analyzed in the mathematical literature \cite{ashwin1992dynamics,golubitsky2003symmetry,aguiar2018synchronization}. Notably, the coupled cell formalism is not limited to dyadic interactions. 
It is formulated in terms of the full dynamical input into each node and can be used to predict admissible patterns of synchrony including those with phase shifts in coupled dynamical systems \cite{golubitsky2016rigid}. The hypergraph treatment instead breaks down the dynamics by order of interaction 
(e.g., dyadic and triadic), as in \cref{eq: dynamics}, for which 
the approach presented here is more readily applicable  than that of the coupled cell formulation (see \cref{app: coupled cell}). }






\add{\textit{Background:}} First, we define the general form of the dynamics on hypergraphs that is being considered. 
A hypergraph consists of a set of $N$ nodes and a set of hyperedges $e_j\in\mathcal{E}$. In this work, we focus on undirected hyperedges. 
Let $\mathcal{E}_i\subset \mathcal{E}$ be the set of hyperedges 
that contain node $i$. Each hyperedge $e_j\in\mathcal{E}_i$ contains a set of nodes $e_j=\{i,j_1,...,j_{m-1}\}$. The order of the hyperedge $e_j$
is $m$, which is the number of nodes including $i$ that are part of it. Thus, $m=2$ corresponds to dyadic edges \add{(between pairs of nodes)}, $m=3$ to triadic edges \add{(between triples of nodes)}, etc.

The adjacency structure can be defined in terms of the collection of $m$ incidence matrices $I^{(m)}$, one for each edge order $m$ (as illustrated in Fig.3 of Ref.\cite{battiston2020networks}). 
Let $\mathcal{E}^{(m)}_i$ be the set of hyperedges of order $m$ containing the node $i$. Then, for the simplest case of homogeneous edge coupling 
the nonzero elements of  the incidence matrix are $[I^{(m)}]_{i,e}=1$ if $e\in\mathcal{E}^{(m)}_i$ for each order $m$.
 
The state of each node can be expressed as an $n$-dimensional real valued vector $\bm x_i\in \mathbb{R}^n$ whose evolution is as follows:
\begin{align}\label{eq: dynamics}
	\dot{\bm x}_i=&\bm F(\bm x_i) + \sum\limits_{m=2}^{d}\sigma^{(m)}\sum\limits_{e\in \mathcal{E}^{(m)}} [I^{(m)}]_{i,e} \bm G^{(m)}(\bm x_i,\bm x_{e\backslash i}),
\end{align}
where $d$ is the maximum edge order present in the hypergraph. 
Here, $\sigma^{(m)}$ denotes the strength of the $m$th order coupling. 

The function $F(\bm x_i)$ describes the internal dynamics of the node $i$, and $G(\bm x_i,\bm x_{e\backslash i})$ is a coupling function corresponding to the influence of the hyperedge $e$ 
on node $i$, where $\bm x_i$ is the state of the node $i$, and $\bm x_{e\backslash i}$ is the state of the rest of the edge. 
\add{This setup includes the specific case when the interaction hypergraph is a simplicial complex where the additional requirement that each subset of nodes in a hyperedge forms a hyperedge of lower order must be satisfied. }

Here we focus on noninvasive coupling functions to keep notation minimal (see the companion paper \cite{salova2021analyzing} for more general coupling). 
\add{Specifically, we assume that the non-dyadic coupling functions} for edges of order $m$ are of the form $\bm G^{(m)}\left(\sum\limits_{l=1}^{m-1} \bm x_{j_l} - (m-1)\bm x_i \right)$. \add{We refer to this type of coupling as Laplacian-like and note that we assume that for each $m$, $\bm G^{(m)}$ can not be reduced to the sum of lower-order interactions.} Coupling functions of this form are natural, for instance, for higher order networks of phase oscillators \cite{skardal2020memory,skardal2019abrupt}, and are not limited to systems with one-dimensional node states.

\begin{figure}
	\includegraphics[scale=.7]{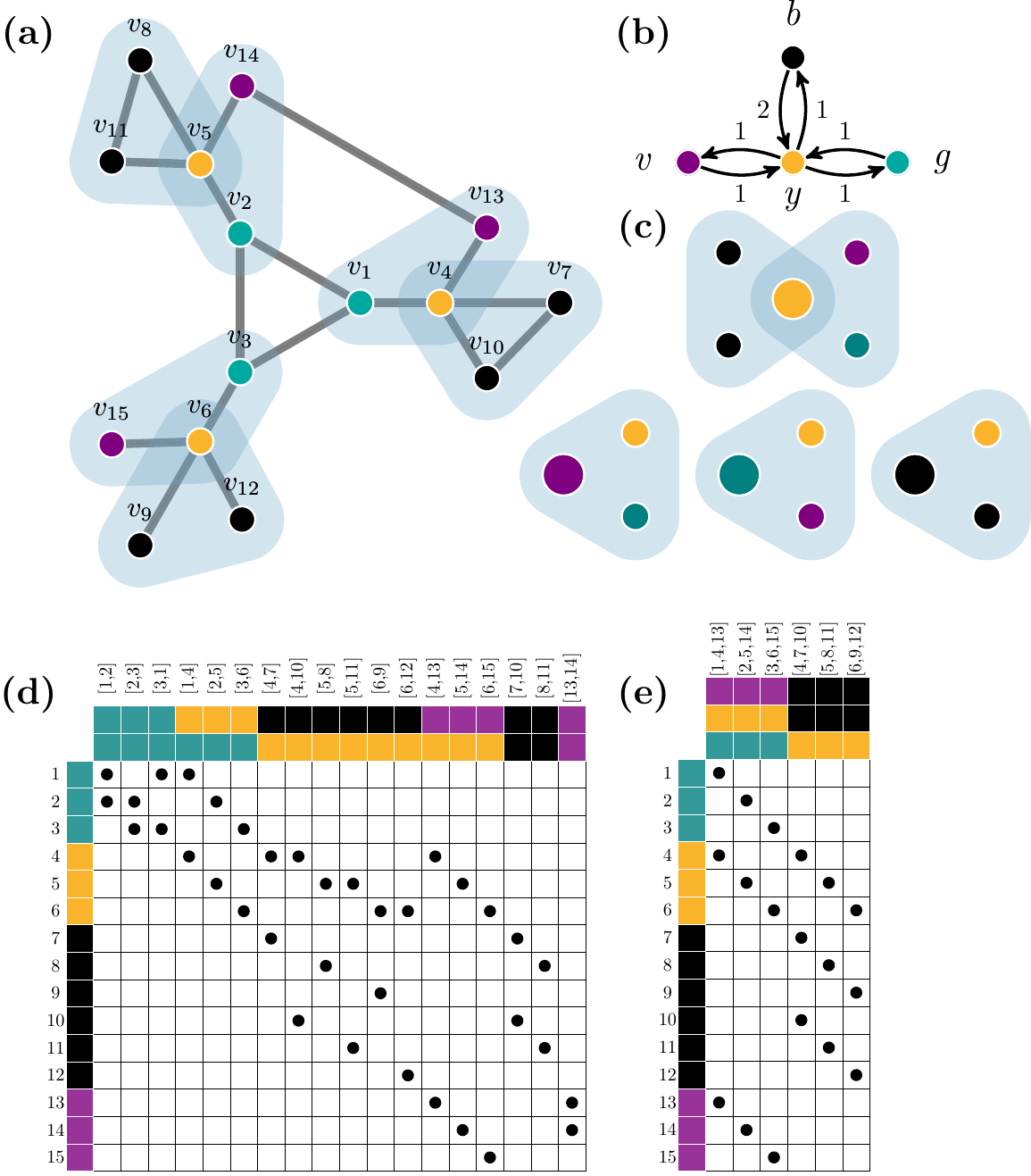}
	\caption{(a) Admissible pattern of synchronization into four node clusters  
	on the specified hypergraph. (b) The dyadic \add{effective interactions}. (c) The triadic \add{effective interactions}, with 
	nodes undergoing that effective dynamics shown in larger size. (d) Incidence matrix for dyadic interactions, $I^{(2)}$. Dots represent ones. Row label colors represent the node clusters, column label colors represent the edge clusters. There are 6 types of dyadic edge clusters. (e) Incidence matrix for triadic interactions, $I^{(3)}$. There are two types of triadic edge clusters.
	}
	\label{fig: clusters}
\end{figure}

\add{\textit{Patterns of cluster synchronization:}} Cluster synchronization is  manifested by groups of nodes following the same trajectory over time, $\bm x_{i_1}(t)=...=\bm x_{i_L}(t)$, where the groups are not fully synchronized with one another. We call each group of synchronized nodes a ``node cluster" and assuming $K$ distinct groups exist, denote them by $C_1, C_2 \dots, C_K$. \add{We refer to the assignment of nodes into clusters as ``patterns of cluster synchronization''}. 
\add{An example of a 4-node cluster pattern 
is shown in Fig.~\ref{fig: clusters}(a).}
The set of dynamic trajectories followed by the nodes in each cluster, the ``node cluster trajectories",
can be expressed as $\bm s_1(t),...,\bm s_K(t)$. For compactness we make the time-dependency
implicit and use the notation $\bm s_1,...,\bm s_K$. 

Likewise, we consider ``edge clusters" and ``edge cluster trajectories". A hyperedge of order $m$ can be characterized by the node clusters to which the $m$ nodes it connects together belong.  (We only need to consider the unordered set if the edges are undirected.) 
All the hyperedges of order $m$ that couple together the same set of node clusters constitute an edge cluster. Assuming $K_m$ distinct edge clusters exist for each order $m$, they are denoted by $C_1^{(m)}, C_2^{(m)}, \dots, C_{K_m}^{(m)}$. The edge cluster trajectories are denoted by $\bm s_{C_1^{(m)}}, \bm s_{C_2^{(m)}}, \dots, \bm s_{C_{K_m}^{(m)}}$,
where $\bm s_{C_j^{(m)}}$ is the set of dynamic trajectories followed by the nodes involved in the $j$th edge cluster.
The node and edge clusters with their corresponding trajectories will be used to facilitate stability calculations.

\add{{\it Admissible patterns:}} For 
networks with purely dyadic interactions, 
equitable partitions can be used to determine the admissible synchronized clusters \cite{kamei2013computation,schaub2016graph}
as well as other patterns of synchronization \cite{salova2020decoupledPRR}.
Equitable partitions divide the network into cells, where each node in a cell $C_i$ receives the same input from any cell $C_j$ including the nodes within its own cell, $i=j$. Each cell thus 
defines a cluster of nodes that could be synchronized. In case of noninvasive coupling, 
the conditions above only have to hold for $i\neq j$ (in which case the partition is called an external equitable partition), since the terms representing the effect of nodes within the same cluster upon one another becomes zero for that partitioning. 

The same idea holds for networks with higher order interactions, 
but the partitions need to be defined in terms of interactions of all orders. 
%
%
The incidence matrices $I^{(m)}$ for the system can be used to obtain explicit partitions 
into non-overlapping cells of node clusters and edge clusters. 
\add{For example, the 4-node cluster state shown~\cref{fig: clusters}(a) has corresponding $I^{(2)}$ and $I^{(3)}$ shown in~\cref{fig: clusters}(d-e) respectively. }
%
The partition is equitable if each node in a given node cluster 
gets the same input from each edge cluster \add{(see \cref{eq: incidence cs})}. 
Many distinct partitions, each one corresponding to a different pattern of cluster synchronization, can be admissible for a given hypergraph. 
\add{How to identify possible admissible partitions based on existing methods developed for coupled cell networks and dyadic networks is discussed in \cref{App: admissible}.}


%

Figure \ref{fig: clusters}(a) shows a 4-node 
cluster synchronization pattern. 
The structure of the hypergraph is an extension of the network shown in Fig.~1 of Ref. \cite{zhang2020symmetry}, with extra hyperedges added to represent the higher order interactions, and 
extra edges
added to highlight that strict symmetry conditions are not necessary for our framework. 
The nodes can be divided into four non-overlapping 
node clusters which we label by their number for convenience in mathematical formulas, $C_1, C_2, C_3, C_4$, or equivalently by their color for convenience when referring to a figure, $C_g, C_y, C_b, C_v$, corresponding to green, yellow, black, and violet. $C_1=C_g=\{1,2,3\}$, $C_2=C_y=\{4,5,6\}$, $C_3=C_b=\{7,8,9,10,11,12\}$, and $C_4=C_v=\{13,14,15\}$. 
With respect to edge clusters, there are 6 distinct dyadic order edge clusters as shown by the identical color combinations in the column labels of \cref{fig: clusters}(d). There are two distinct triadic order edge clusters shown by the identical color combinations in the column labels of \cref{fig: clusters}(e), where $C^{(3)}_1=C^{(3)}_{gyv}=\{[1,4,13],[2,5,14],[3,6,15]\}$ and $C^{(3)}_2=C^{(3)}_{ybb}=\{[4,7,10],[5,8,11],[6,9,12]\}$. These node and edge clusters together form an external equitable partition.
Therefore, this particular partition corresponds to an admissible pattern of synchronization.

Formally, any admissible pattern of cluster synchronization on a hypergraph with Laplacian-like coupling must satisfy the following condition:
\begin{align}\label{eq: incidence cs}
	\sum\limits_{e_j\in C^{(m)}_k} I^{(m)}_{ij} = \sum\limits_{e_j\in C^{(m)}_k} I^{(m)}_{i'j},
\end{align}
for $i,i'\in C_{l}$, 
where we are summing over the $m$-th order hyperedges $e_j$ that are in edge cluster $C_k^{(m)}$, where the terms coming from edge clusters $C_k^{(m)}$ that contain only nodes in $C_l$ can be ignored due to noninvasive coupling.

\add{{\it Effective dynamics:}} The dynamics for an admissible pattern 
can be expressed via a set of \add{\textit{effective incidence matrices}}, $I_{\text{eff}}^{(m)}$, representing the interactions between nodes of different clusters. 
\add{Each $I_{\text{eff}}^{(m)}$ can be formed 
by considering one representative node from each cluster and calculating how many different types of hyperedges of order $m$ it is part of. Hyperedges that contain only nodes in the same cluster should be excluded.
See Ref.\cite{salova2021analyzing} for more details.} 
For 
\cref{fig: clusters}(a)  
these matrices are:
\begin{align}
	I^{(2)}_{\text{eff}}=
\begin{blockarray}{cccc}
	&\begin{turn}{90} $[by]$ \end{turn} & \begin{turn}{90} $[yv]$ \end{turn} & \begin{turn}{90} $[yg]$ \end{turn} \\
	\begin{block}{c(ccc)}
		b~~&	1 &   &   \\
		y~~&	2 & 1 & 1 \\
		v~~&	  & 1 &  \\
		g~~&	  &   & 1 \\
	\end{block}
\end{blockarray}~~,~~~
I^{(3)}_{\text{eff}}=
\begin{blockarray}{ccc}
	&\begin{turn}{90} $[bby]$ \end{turn} & \begin{turn}{90} $[gyv]$ \end{turn} \\
	\begin{block}{c(cc)}
		b~~&	1 &      \\
		y~~&	1 & 1 \\
		v~~&	  & 1   \\
		g~~&	  & 1   \\
	\end{block}
\end{blockarray}~~.
\end{align}
\vspace{-1.3em}\\
The \add{graphical representation of these matrices} (illustrated in \cref{fig: clusters} (b-c)) can be used to read out the time evolution of each node. For instance, every node in the yellow cluster evolves according to:
\begin{align}
\dot{ \bm x}_y &= \bm F(\bm x_y) +\bm G^{(2)}(\bm x_g-\bm x_y)\nonumber\\&+\bm G^{(2)}(\bm x_v-\bm x_y)+2\bm G^{(2)}(\bm x_b-\bm x_y)\nonumber\\&+\bm G^{(3)}(\bm x_b+\bm x_b-2\bm x_y)+\bm G^{(3)}(\bm x_v+\bm x_g-2\bm x_y),
\end{align}
with analogous equations describing the time evolution for each distinct cluster. 
Here, $G^{(m)}$ is expressed taking 
into account the Laplacian-like coupling assumption. 

\add{\textit{Stability:}} For dyadic networks, stability analysis of cluster synchronization patterns is well developed \cite{pecora2014cluster,cho2017stable}. However, in the presence of higher order coupling, the Jacobian acquires additional terms. Here, we show how the Jacobian can be block diagonalized by using the incidence matrices for a given cluster synchronization pattern, thus \add{simplifying} stability calculations for general dynamics on hypergraphs.

First, we define a Laplacian corresponding to the $k$th edge cluster 
of order $m$ as follows:
\begin{align}\label{eq: Laplacian}
\mathcal{L}^{(m)}_{k}&=m\mathcal{D}^{(m)}_{k}-I^{(m)}_k[I^{(m)}_k]^T,
\end{align}
where $I^{(m)}_k$ is an $N\times |C^{(m)}_k|$ matrix consisting of the columns of $I^{(m)}$ that correspond to the hyperedges in the $k$th cluster of order $m$ (here, $|C^{(m)}_k|$ denotes the number of unique elements in the edge cluster $C^{(m)}_k$). For instance, for the $C^{(3)}_2$ edge cluster 
of the hypergraph in \cref{fig: clusters}(a), $I^{(3)}_2$ is obtained by keeping the last $3$ columns of $I^{(3)}$ in \cref{fig: clusters}(e). Additionally, $\mathcal D^{(m)}_{k}$ is a diagonal matrix with elements $[\mathcal D^{(m)}_{k}]_{ii}$ corresponding to the number of $m$th order edges in the $k$th edge cluster node $i$ is part of ($[\mathcal D_k^{(m)}]_{ii}=\sum\limits_{j=1}^N [I^{(m)}_k]_{ij}$). The form of the projected Laplacians (the $\mathcal{L}^{(m)}_k$ matrices) is similar to that of the generalized Laplacian \cite{lucas2020multi}.
\add{For each projected Laplacian matrix $\mathcal{L}^{(m)}_{k}$, we can define a corresponding projected adjacency matrix with diagonal elements equal to zero:}
\begin{align}\label{eq: adjacency}
	\mathcal{A}^{(m)}_{k}=(m-1)\mathcal{D}^{(m)}_{k}-\mathcal{L}^{(m)}_{k}	
\end{align} 

The variational equation for linear stability depends on all node clusters and all edge clusters of all orders. For a specific pattern of cluster synchronization, it is:
\begin{align}\label{eq: lap stab}
{\delta \dot{\bm x}}&=
\bigg(\sum\limits_{k=1}^K E_k \otimes J\bm F(\bm s_k)
-\sum\limits_{m=2}^d\sigma^{(m)}\cdot\\
&\Big(\sum\limits_{k=1}^{K_m}\sum\limits_{l\in \{C_k^{(m)}\}} E_{l} \mathcal{L}^{(m)}_k\otimes J\bm G^{(m)}(\bm s_l, \bm s_{C^{(m)}_k\backslash l})\Big)\bigg)\delta\bm  x,\nonumber
\end{align}
where $E_k$ denotes the diagonal cluster indicator matrix encoding which nodes are in cluster $C_k$ ($[E_k]_{ii}=1$ if $i\in C_k$ and $[E_k]_{ii}=0$ otherwise). Additionally, $\{C_k^{(m)}\}$ is a set of \textit{unique} node clusters included in the $k$th edge cluster, (e.g., in Fig.~\ref{fig: clusters}, $\{C^{(3)}_{ybb}\}=\{y,b\}$). 
Finally, $\bm s_{C^{(m)}_k\backslash l}$ is the set of all the trajectories of nodes included in edge cluster $C_k^{(m)}$, excluding those nodes in node cluster $l$.
The partial derivatives are computed as:
\begin{align}\label{eq: jg}
J\bm G^{(m)}&(\bm s_l, \bm s_{C^{(m)}_k\backslash l})_{p,q}\nonumber \\&=\dfrac{\partial \bm G^{(m)}_p\left(\sum\limits_{j=1}^{m-1} \bm x_{j}-(m-1)\bm x_0\right)}{\partial  [\bm x_{2}]_q}\bigg|_{\substack{\bm x_0=\bm s_l,\\\bm x_{j}= [\bm s_{C^{(m)}_k\backslash l}]_j}}\nonumber \\&=\dfrac{\partial \bm G_p^{(m)}(z)}{\partial \bm z_q}\bigg|_{\bm z=\sum\limits_{j=1}^{m-1}\bm [\bm s_{C^{(m)}_k\backslash l}]_j-(m-1)\bm s_l}
\end{align}
where $[\bm s_{C^{(m)}_k\backslash l}]_j$ is the $j$th trajectory in the set $\bm s_{C^{(m)}_k\backslash l}$.

The key implication of \cref{eq: lap stab,eq: jg} is that to block diagonalize the Jacobian for the entire Laplacian-like coupled system, it is sufficient to simultaneously block diagonalize the following matrices:
\begin{align}\label{eq: mats Laplacian}
\{E_1,...,E_K,\mathcal{L}^{(2)},\mathcal{L}^{(3)}_{1},...,\mathcal{L}^{(3)}_{K_3},...,\mathcal{L}^{(d)}_{1},...,\mathcal{L}^{(d)}_{K_d}\}.
\end{align}
These are the indicator matrices, $E_j$, and the $\mathcal{L}^{(m)}_k$ matrices for a specific pattern of cluster synchronization.
\begin{figure}
	\includegraphics[scale=.4]{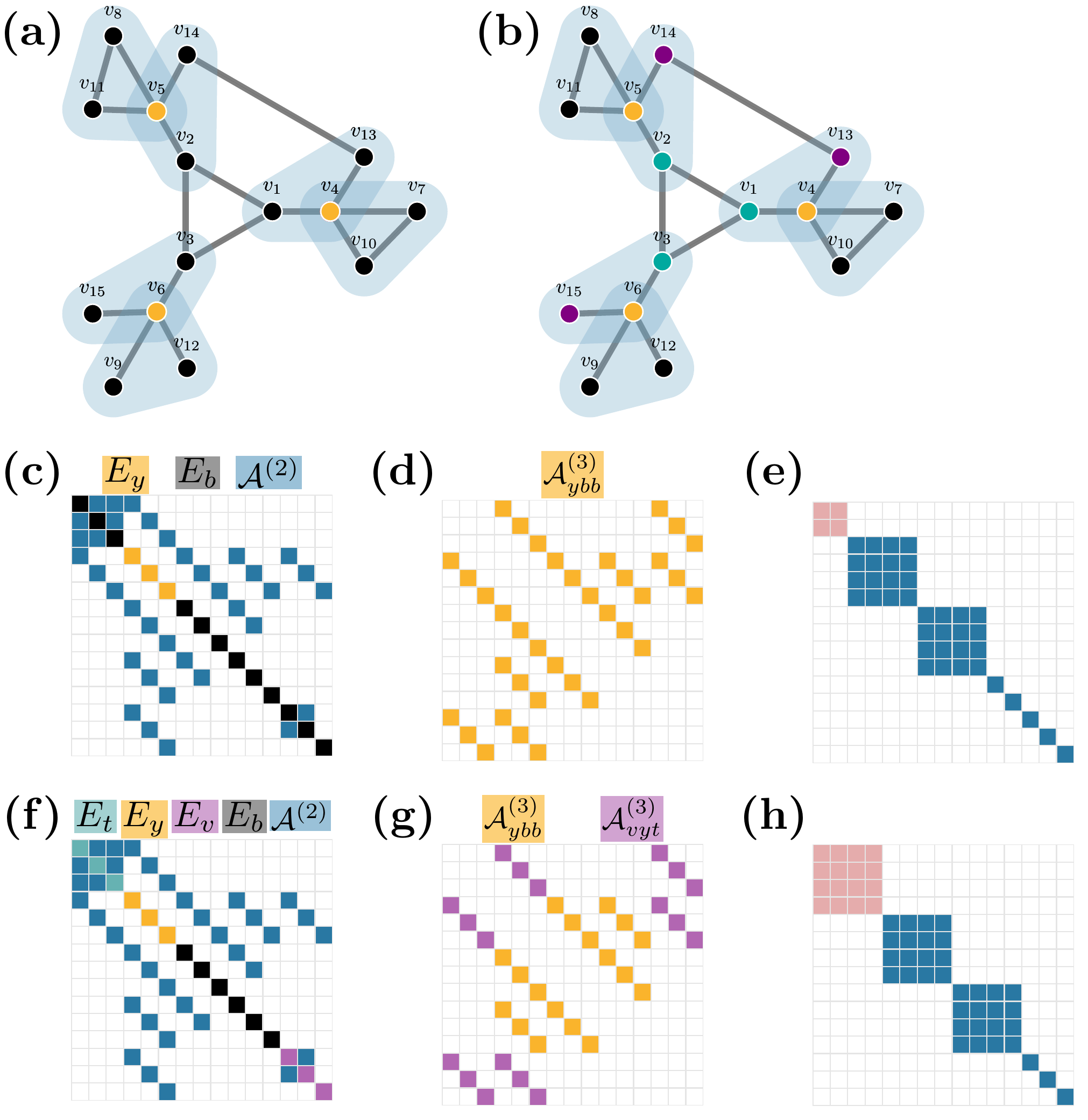}
	\caption{Admissible states with (a) two and (b) four node clusters. \add{In (c-h) colored squares represent ones, white represent zeros.} For the two cluster state, the set of matrices \add{(\cref{eq: mats Laplacian})}  representing (c) dyadic and (d) triadic interactions 
	that have to be simultaneously block diagonalized to (e) block diagonalize the Jacobian (pink corresponds to parallel and blue to transverse perturbations).  For the four cluster state, the set of matrices representing (f) dyadic and (g) triadic interactions and (h) the resulting block diagonalized Jacobian. 
	\add{For visualization purposes the projected adjacency matrices, rather then the projected Laplacian matrices are shown, see \cref{eq: adjacency}. }
	}
	\label{fig: mats}
\end{figure}

\add{Two different examples of this procedure are 
shown \cref{fig: mats} 
using the simultaneous block diagonalization algorithm from Ref.\cite{zhang2020symmetry}.
Figure~\ref{fig: mats}(a) shows a 
two-cluster and Fig~\ref{fig: mats}(b) a four-cluster state. For the two-cluster state, since each node participates in one unique triadic edge pattern, it is sufficient to simultaneously block diagonalize the cluster indicator matrices ($E_y$ and $E_b$), the dyadic Laplacian $\mathcal{L}^{(2)}$, and the projected triadic Laplacian $\mathcal{L}^{(3)}$ as illustrated in \cref{fig: mats}(c-e) where, for the sake of visualization, the related adjacency matrices $\mathcal{A}^{(2)}$ and $\mathcal{A}^{(3)}$ (\cref{eq: adjacency}) 
are shown.  
In the case of the four-cluster state, there are two triadic edge clusters so two matrices
representing triadic patterns, $\mathcal{L}^{(3)}_1$ and $\mathcal{L}^{(3)}_2$ (with corresponding $\mathcal{A}^{(3)}_1$ and $\mathcal{A}^{(3)}_2$ shown in \cref{fig: mats}(g)), must be included in the simultaneous block diagonalization with the full set of matrices shown in \cref{fig: mats}(f-h)).
}

\begin{figure*}
	\includegraphics[scale=.55]{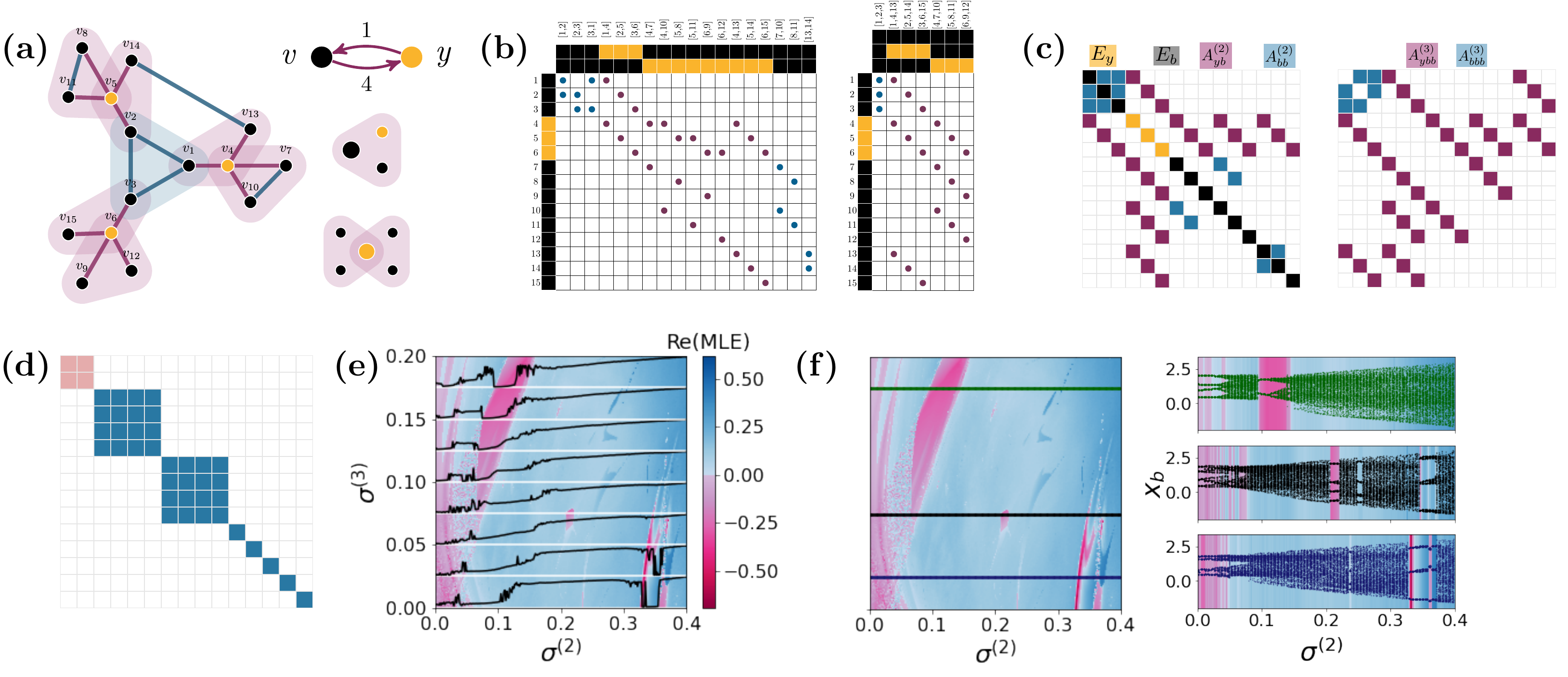}
	\caption{Linear stability calculation for the system discussed in \cref{Eq: dyn individual,Eq: dyn coupling} performed using the accompanying code \cite{salova2021code}. (a) Left: a hypergraph with 
		attractive (blue) and repulsive (violet) coupling (\cref{Eq: dyn coupling}) and with nodes that obey the same dynamical equation (\cref{Eq: dyn individual}). Right: the \add{representation of effective dyadic and triadic interactions}. (b) Structure of $I^{(2)}$ (left) and $I^{(3)}$ (right). Black and yellow represent the distinct clusters, blue and violet represent the distinct coupling types. (c) Matrices used in simultaneous block diagonalization to perform the stability analysis. (d) Jacobian structure after block diagonalization. Pink represents parallel and blue transverse perturbations. (e) Linear stability diagram for a fixed parameter $\beta=1.8$ and various values of dyadic and triadic coupling strengths, $\sigma^{(2)}$ and $\sigma^{(3)}$. Pink areas are linearly stable, blue areas are not linearly stable. Black lines correspond to direct simulation of standard deviations from the average cluster trajectory for each of the $\sigma^{(3)}$ values in white. (f) Left: stability diagram with three distinct $\sigma^{(3)}$ values shown with different colored solid lines. Right: bifurcation diagram for the three distinct $\sigma^{(3)}$ values shown in the corresponding color. Horizontal axis represents the dyadic coupling strength, vertical axis corresponds to the states of black nodes $x_{black}$ in the past $100$ time steps. Background colors represent the calculated linear stability for each value of $\sigma^{(2)}$.}
	\label{fig: stab}
\end{figure*}

The results 
for binary hyperedges generalize to systems with different types of nodes and edges \cite{della2020symmetries}, and thus are applicable to multilayered networks \add{with higher order interactions}\add{, as demonstrated in \cref{App: different node edge types}. Specifically, nodes of the same type can synchronize if they receive the same input from all interaction orders and edge types within each order, and Jacobian block diagonalization can be obtained by simultaneously block diagonalizing the set of matrices in \cref{eq: mats Laplacian different}.}


An example hypergraph with different types of hyperedges is shown in \cref{fig: stab}(a), where distinct colors (blue and violet) illustrate distinct hyperedge types. 
These are also highlighted in \cref{fig: stab}(b-c) with different colors corresponding to different edge types in 
the labeled incidence matrices. We consider a two-node cluster state on this hypergraph and use  black and yellow colors to distinguish nodes in each cluster.  For this state, \cref{eq: incidence cs} establishing the condition for cluster synchronization holds for all types of edges and all coupling orders. 

In order to obtain concrete linear stability results, we need to impose specific dynamical equations to describe the evolution of the system. We use the optoelectric oscillator dynamics used in experiments in Ref.\cite{hart2017experiments}, with one-dimensional discrete time node dynamics \begin{align}\label{Eq: dyn individual}
	F(x_i)=\beta \sin^2(x_i+\pi/4)
\end{align} 
and coupling functions
\begin{align}\label{Eq: dyn coupling}
	G^{(2)}(x_i, x_j)=&\text{sgn}_{ij}\sigma^{(2)}[F(x_j)-F(x_i)],\\ G^{(3)}(x_i,x_j,x_k)=&\text{sgn}_{ijk}\sigma^{(3)}\sin(x_i+x_j-2x_k).\nonumber
\end{align} 
\add{We pick Laplacian coupling for dyadic interactions and Laplacian-like coupling for triadic interactions. Thus, if $\sigma^{(3)}=0$, our equations reduce to the dynamics in Refs.\cite{hart2017experiments,zhang2020unified,zhang2020symmetry}.} \add{Here, $\text{sgn}_{ij}=1$ if the coupling on the edge between $i$ and $j$ is attractive (shown in blue in \cref{fig: stab}) and $\text{sgn}_{ij}=-1$ if the coupling is repulsive (shown in violet in \cref{fig: stab}). Similarly, $\text{sgn}_{ijk}=1$ for attractive (blue) hyperedges, and $\text{sgn}_{ijk}=-1$ for repulsive (violet) hyperedges.}

To avoid complications from multistability, we analyze a two cluster state and 
make edges connecting only nodes that are in the same cluster attractive and all other edges repulsive.
Keeping 
$\beta$ constant, we vary $\sigma^{(2)}$ and $\sigma^{(3)}$ to determine the linear stability regions for different parameter regimes. 
The analysis is shown in \cref{fig: stab}, with more details in \cref{app: BD details}.
Figure~\ref{fig: stab}(a-b) shows the analogous plots to Fig.~\ref{fig: clusters}(a-e) with the state, the \add{representation of effective interactions between nodes}, and incidence matrices respectively. Figure~\ref{fig: stab}(c) shows 
the set of matrices that need to be simultaneously block diagonalized with the resulting Jacobian in Fig.~\ref{fig: stab}(d).  Figure~\ref{fig: stab}(e) is the linear stability plot 
demonstrating sensitive dependence on both 
$\sigma^{(2)}$ and $\sigma^{(3)}$, with the changes in the stability properties of the system showing correspondence to different regions of the bifurcation diagram shown in Fig.~\ref{fig: stab}(f). 

\add{\textit{Conclusion:}} Systems of dynamical elements coupled on hypergraphs can show intricate synchronization patterns beyond full synchronization.
A crucial aspect of 
understanding such systems is their stability analysis. 
We show how to use the structure of the incidence matrices to determine the admissibility of cluster synchronization patterns
and reduce the dimension of stability calculations.
Our formulation is in terms of node clusters and the hyperedge clusters that are induced by
the synchronization pattern of the entire set of nodes coupled on each hyperedge.
This provides a general way to organize the analysis of dynamical processes on hypergraphs.

Unlike previous work, our analysis is not restricted to dyadic interactions, full synchronization, or non-intertwined clusters.
Our results open up new opportunities for detailed analysis of systems of theoretical and practical significance, as well as investigating the role of higher order interactions in stabilizing or destabilizing different states.


\textit{Acknowledgments.}
The authors thank Yuanzhao Zhang and Adilson Motter for helpful discussions about theory and code.

\bibliographystyle{unsrt}
\bibliography{biblio_hypergraphs}

\appendix

\section{Finding admissible patterns of cluster synchronization}\label{App: admissible}

Here, we discuss how to obtain the admissible patterns of cluster synchronization in systems with higher order interactions. To obtain these patterns, we need to search for partitions in which each node of the same type gets the same dynamical input from all the other nodes. The cells of such partitions determine which nodes can be fully synchronized. Mathematically, these conditions for systems described in \cref{eq: dynamics} are contained in \cref{eq: incidence cs}. 

While it is easy to use \cref{eq: incidence cs} to check whether a given pattern of synchronization is admissible for a specific hypergraph structure, it is more challenging to identify all of the possible admissible patterns. 
Here we show how methods developed for coupled cell networks and systems with dyadic interactions (e.g., Refs.\cite{kamei2013computation,sorrentino2016complete,pecora2014cluster}) can be modified for finding the cluster synchronization patterns for higher order interactions. 
We demonstrate this for undirected hypergraphs with Laplacian-like coupling. In Ref.\cite{salova2021analyzing}, we demonstrate how to obtain such patterns for systems without the Laplacian-like coupling constraint (which we refer to as adjacency coupling).
%
%
The main difference between Laplacian and adjacency coupling is that the former relaxes some of the conditions on state admissibility. Namely, for Laplacian coupling, we can ignore the contributions from hyperedges on which all the nodes belong to the same node cluster. We outline two approaches for obtaining the admissible cluster synchronization states for Laplacian-like coupling.

\begin{figure}	
	\includegraphics[scale=.45]{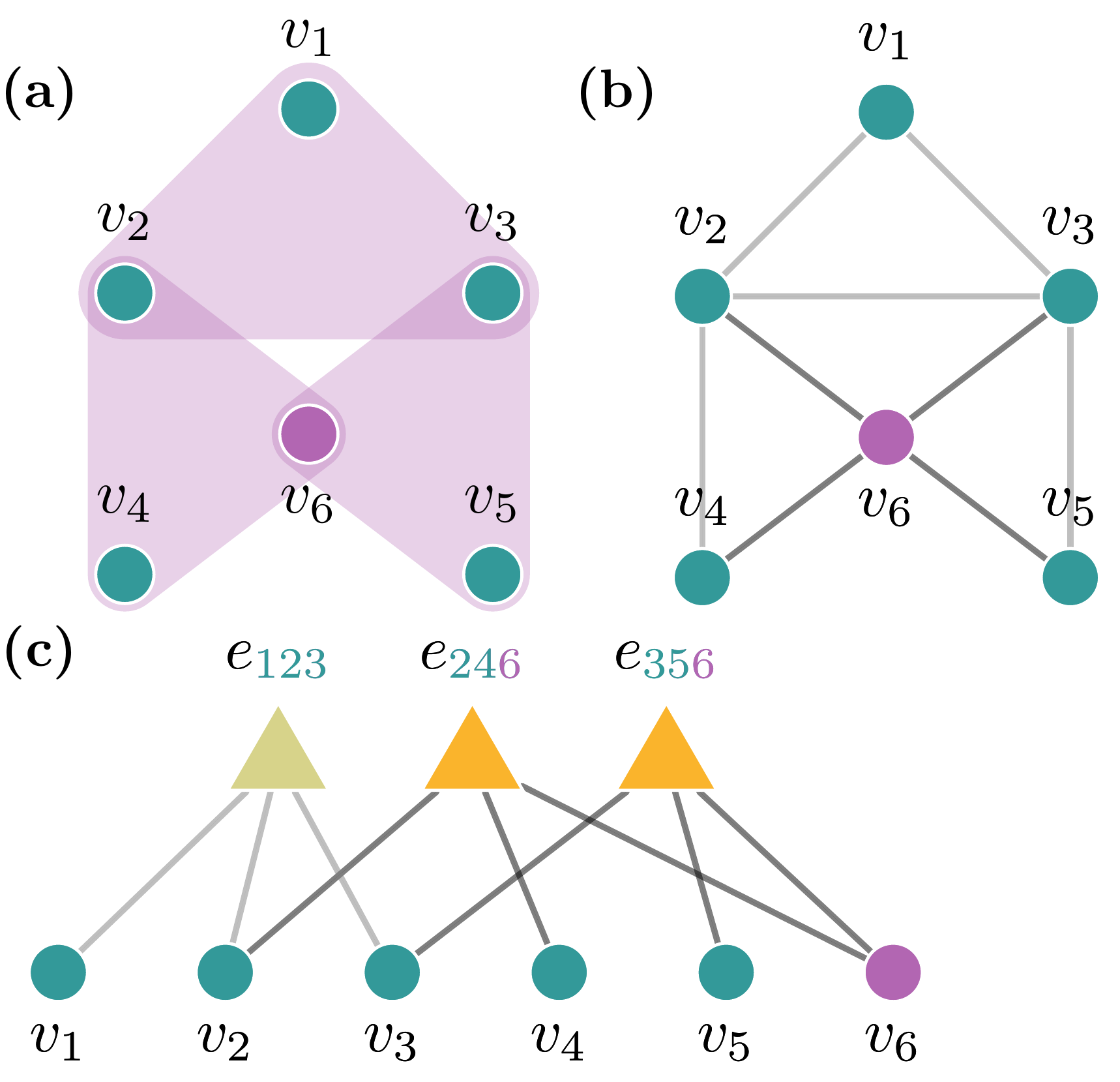}
	\caption{
Identifying admissible patterns.  (a) Shown is an admissible two-cluster state on a hypergraph with three triadic hyperedges. (Distinct node clusters are shown with distinct colors.) We can identify the two-cluster state as a possible admissible pattern using two different approaches: (b) the projected network and (c) the bipartite representation. We then need to verify whether the identified pattern is admissible on the actual hypergraph using \cref{eq: incidence cs}. 
	(b) The projected network with edges containing the nodes in the same cluster 
	shown in lighter gray.  Existing methods \cite{kamei2013computation} allow us to identify that the two-cluster state is admissible on this projected network. 
(c) The bipartite representation of the hypergraph with hyperedges and nodes represented by triangles and circles respectively. Existing methods~\cite{kamei2013computation} allow us to identify that the two-cluster state is admissible. 
Note, edges representing node membership in a hyperedge that only contains 
teal nodes do not affect the dynamics of the cluster state, therefore are shown in lighter grey.}
	\label{fig: admissible}
\end{figure}

The first way is to consider the \textit{projected networks} (e.g., each higher order triadic edge becomes three dyadic edges between the nodes comprising that hyperedge). The 
admissible patterns of synchronization can be determined for the projected network for all orders of interaction using existing methods \cite{kamei2013computation,sorrentino2016complete,pecora2014cluster}
and then one can check their admissibility on the original hypergraph using \cref{eq: incidence cs}. Consider, for instance, the example in 
\cref{fig: admissible}(a) which demonstrates a hypergraph (containing only triadic edges for simplicity).
The projected network is shown in \cref{fig: admissible}(b) with the edges that can be ignored in the case of Laplacian coupling shown in light grey. 
The existing methods identify that the two-cluster pattern is admissible on the dyadic projection. Then, we can check whether the state is also admissible on the original hypergraph shown in \cref{fig: admissible}(a) using \cref{eq: incidence cs}.

A second way to obtain the admissible patterns of cluster synchronization on hypergraphs is by representing the hypergraph structure as a bipartite network, with two sets of vertices corresponding to nodes and hyperedges respectively (e.g., as demonstrated in \cref{fig: admissible}(c)). Then, the problem can be addressed from the perspective of finding (external) equitable partitions of a network with different types of nodes using the tools from Ref.\cite{kamei2013computation}. As discussed in detail in Ref.\cite{salova2021analyzing}, an extra step to obtain valid cluster synchronization patterns is picking the partitions where the edge partition is induced by the node partition. Note that in \cref{fig: admissible}(c), the light gray edges correspond to the node membership in the edge containing the nodes in the same cluster 
(shown in teal), and can therefore be ignored in the admissibility analysis.

\begin{figure*}[t]
	\includegraphics[scale=1.09]{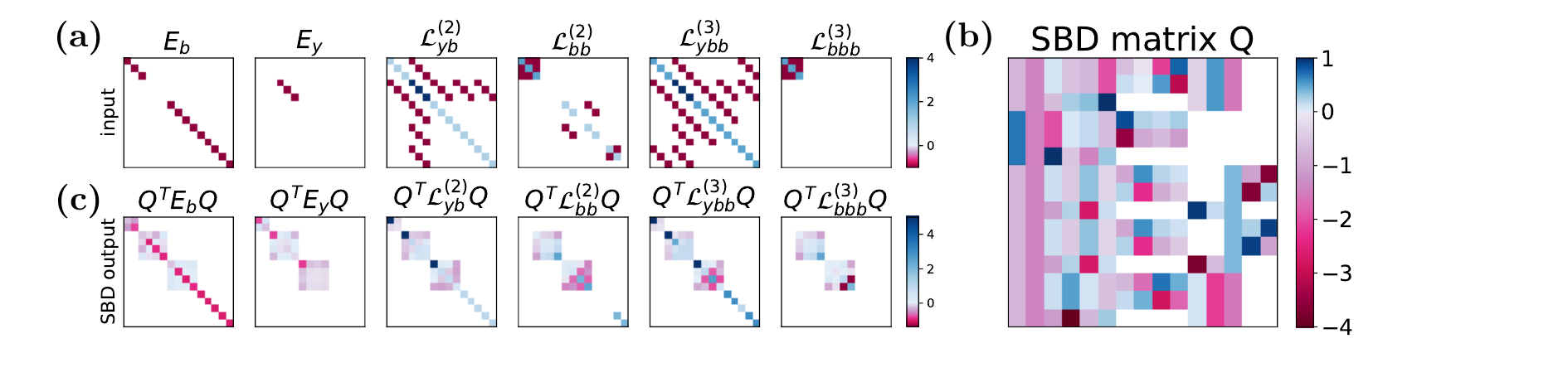}
	\caption{Details of the simultaneous block diagonalization (SBD) procedure summarized in \cref{fig: stab}(c) and (d) of the main manuscript.
	(a) The set of matrices that need to be simultaneously block diagonalized. (b) The SBD matrix $Q$ obtained using the code from Ref.\cite{zhang2020symmetry}. (c) The block diagonalized matrices obtained by transforming the matrices in (a) using $Q$.}
	\label{fig: SBD steps}
\end{figure*}

\section{Systems with multiple node and edge types.}\label{App: different node edge types}

Interactions between nodes of different types via different types of dyadic edges generalize networks with a single node and edge type and can be studied in context of multilayer networks
\cite{kivela2014multilayer,della2020symmetries}.
In the mathematical literature, patterns of synchronization in very general systems with different nodes and edge types are considered in Refs.\cite{golubitsky2005patterns,golubitsky2016rigid}. Here, we specifically consider hypergraphs and generalize the admissibility and stability analysis of cluster synchronization on hypergraphs to systems with different types of nodes and edges.

First, we define a set of relevant incidence matrices $I^{(m)}_{\alpha}$, where $m=2,3,...,d$ corresponds to the order of interactions, and $\alpha=1,...,\kappa_m$ indexes over the edge types for each interaction order $m$. Similarly, we can define $I^{(1)}_{\alpha}$, the ``self-incidence matrix'' for each node type. Then, the evolution of the dynamics of nodes on a hypergraph with different node and edge types can be expressed as:
\begin{align}\label{eq: dynamics 2}
	&\dot{\bm x}_i=\sum\limits_{\alpha=1}^{\kappa_1}[I^{(1)}_{\alpha}]_i\bm F_{\alpha}(\bm x_i) \\&+ \sum\limits_{m=2}^{d}
	\sum\limits_{\alpha=1}^{\kappa_m}
	\sigma^{(m)}_{\alpha}\sum\limits_{e\in \mathcal{E}^{(m)}} [I^{(m)}_{\alpha}]_{i,e} \bm G^{(m)}_{\alpha}(\bm x_i,\bm x_{e\backslash i}), \nonumber
\end{align}
where $\bm F_{\alpha}$ and $\bm G^{(m)}_{\alpha}$ represent the dynamics of the $\alpha$th node or edge type, and $\sigma_{\alpha}^{(m)}$ is the coupling strength for $m$th order edges of different types.

Let $C^{(m)}_{k,\alpha}$ stand for edge clusters of order $m$ and interaction type $\alpha$.
Additionally, we set $C^{(1)}_{k,\alpha}\equiv C_{k,\alpha}$.
Then, the cluster synchronization condition, 
which is very similar to the one introduced in \cref{eq: incidence cs}, can be defined as:
\begin{align}\label{eq: incidence cs different}
	\sum\limits_{e_j\in C^{(m)}_{k,\alpha}} [I^{(m)}_{\alpha}]_{ij} = \sum\limits_{e_j\in C^{(m)}_{k,\alpha}} [I^{(m)}_{\alpha}]_{i'j},
\end{align}
with the equation having to hold for all interaction orders $m$ and interaction types $\alpha$.

The variational equation used for stability analysis is
\begin{align}\label{eq: lap stab 2}
	{\delta \dot{\bm x}}&=
	\bigg(\sum\limits_{\alpha=1}^{\kappa_1}\sum\limits_{k=1}^{K_{\alpha}} [E_{\alpha}]_k \otimes J\bm F_{\alpha}(\bm s_{\alpha,k})
	-\sum\limits_{m=2}^d\sum\limits_{\alpha=1}^{\kappa_m}\sigma^{(m)}_{\alpha}\cdot\\
	&\Big(\sum\limits_{k=1}^{K_{m,\alpha}}\sum\limits_{l\in \{C_{k,\alpha}^{(m)}\}} [E_{\alpha}]_{l} \mathcal[\mathcal{L}^{(m)}_{\alpha}]_k\otimes J\bm G^{(m)}_{\alpha}(\bm s_{\alpha,l}, \bm s_{C^{(m)}_{k,\alpha}\backslash l})\Big)\bigg)\delta \bm x,\nonumber
\end{align}
where the notation is similar to \cref{eq: lap stab}, and the new subscript $\alpha$ indicates the $\alpha$th type of node or interaction type and dynamics. E.g., $\bm s_{\alpha,k}$ stands for trajectory of a $k$th node cluster consisting of nodes of type $\alpha$, and $\bm s_{C_{k,\alpha}^{(m)}}$ is the trajectory of the $k$th $m$th order edge cluster with interactions of type $\alpha$.

In the main body of the manuscript, we present the set of matrices that need to be simultaneously block diagonalized to reduce the dimension of cluster synchronization calculations. Here, we explicitly state what matrices need to be block diagonalized if a hypergraph has multiple node and edge types:
\begin{align}\label{eq: mats Laplacian different}
	\Big\{&\bigcup_{\substack{\alpha=1,...,\kappa\\k=1,...,K_{\alpha}}}[E_{\alpha}]_k,\bigcup_{\alpha=1,...,\kappa_2}\mathcal{L}^{(2)}_{\alpha},\nonumber\\&\bigcup_{\substack{\alpha=1,...,\kappa_3\\i=1,...,K_{3,\alpha}}}[\mathcal{L}^{(3)}_{\alpha}]_k,...,
	\bigcup_{\substack{\alpha=1,...,\kappa_d\\k=1,...,K_{d,\alpha}}}[\mathcal{L}^{(d)}_{\alpha}]_k\Big\},
\end{align}
where, $[E_{\alpha}]_i$ is the indicator matrix for the $i$th node cluster formed by the node type $\alpha$, and $[\mathcal L^{(m)}_{\alpha}]_{i}$ stands for the Laplacian formed for the $i$th edge cluster within the $m$th interaction order and edge type $\alpha$.

In the main manuscript, we apply these results to an example system where all the nodes are of the same type and evolve according to the same dynamics, but where  there are two types of dyadic and two types of triadic edges (visualized in different colors in \cref{fig: stab}(a)). 
There, the attractive and repulsive interactions differ only by the sign of the coupling functions.

In addition to hypergraphs, our approach can be directly applied to multilayer networks as defined in Ref.\cite{della2020symmetries}. There, the authors perform the stability analysis for systems with different types of nodes and dyadic edges and cluster synchronization states arising from symmetries. This section demonstrates how to perform the simplification for multilayer networks with higher order interactions and states that are more general than those arising from symmetries.

\section{Details of block diagonalization}\label{app: BD details}

Here, we provide more details of the stability calculation in \cref{fig: stab}, specifically obtaining the block diagonalization of the matrices in 
\cref{fig: stab}(c) that results in the transformed Jacobian structure demonstrated in \cref{fig: stab}(d) with the parallel perturbation block shown in pink and transverse perturbation blocks shown in blue. The calculation was performed using the simultaneous block diagonalization algorithm of Ref.\cite{zhang2020symmetry} as implemented in \cite{salova2021code}.

In \cref{fig: SBD steps}(a), we show the set of matrices that needed to be simultaneously block diagonalized to obtain the block diagonalization of the Jacobian. Using the notation introduced in \cref{App: different node edge types} these are, $E_b=[E_1]_1$, $E_y=[E_2]_1$, $\mathcal{L}_{yb}^{(2)}=\mathcal{L}^{(2)}_{1}$, $\mathcal{L}_{yb}^{(2)}=\mathcal{L}^{(2)}_{2}$, $\mathcal{L}_{ybb}^{(3)}=[\mathcal{L}^{(3)}_{1}]_1$, and $\mathcal{L}_{bbb}^{(3)}=[\mathcal{L}^{(3)}_{2}]_1$. The calculated unitary block diagonalization matrix $Q$ is demonstrated in \cref{fig: SBD steps}(b).

Denoting $\bm \eta = Q^T\bm x$ and rewriting \cref{eq: lap stab} in terms of this new variable, we get
\begin{align}\label{eq: lap stab 2}
	&{{\delta\bm \eta[t+1]}}=
	\bigg(\sum\limits_{k=1}^K Q^TE_k Q \otimes J\bm F(s_k)
	-\sum\limits_{m=2}^d\sigma^{(m)}\cdot\\
	&\Big(\sum\limits_{k=1}^{K_m}\sum\limits_{l\in \{C_k^{(m)}\}} Q^TE_{l} \mathcal{L}^{(m)}_kQ\otimes J\bm G^{(m)}(\bm s_l, \bm s_{C^{(m)}_k\backslash l})\Big)\bigg)\delta\bm \eta[t],\nonumber
\end{align}
where all the terms have the block diagonal form demonstrated in \cref{fig: SBD steps}(c). The magnitude of numerically calculated elements in \cref{fig: SBD steps}(c) shown in white is under $10^{-13}$. The top left $2\times 2$ block of each matrix corresponds to perturbations parallel to the synchronization manifold. The other blocks (two $4\times 4$, five $1\times 1$) correspond to transverse perturbation and are used in calculating the maximum transverse Lyapunov exponent. We used $10^5$ time steps for the stability calculations, and code used for these calculations is available in Ref.\cite{salova2021code}.

\section{Relating to coupled cell network literature}\label{app: coupled cell}

The coupled cell network formalism leads to general results relating network topology to patterns of synchronization (e.g., cluster synchronization and splay states) in systems that can have higher-order interactions beyond dyadic \cite{stewart2003symmetry,golubitsky2005patterns}. Many of the recent advances in that research area are summarized in Ref.\cite{golubitsky2016rigid}. The authors consider systems of coupled ODEs where each variable corresponds to the dynamics of a specific node. In that formalism, instead of an edge corresponding to specific dyadic or higher-order term in the interaction function, the edges 
(most generally, directed) ``specify which variables occur (perhaps nonlinearly) in which components of the ODE, and encode (via
input sets of arrows) when components of the ODE involve
the same function'' \cite{golubitsky2016rigid}. Although this is very general, as we show below, for considering 
dynamics on hypergraphs composed of additive dyadic, triadic, etc. interactions, the approach laid out in this manuscript offers a more direct treatment.

As an example illustrating the difference between representing higher order interactions via a hypergraph with higher-order interactions and via the coupled cell formalism, consider the dynamics on the hypergraph shown in \cref{fig: admissible} with the Laplacian-like coupling assumption. In the notation of this manuscript,
\begin{align}\label{eq: node formalism}
	\dot{x}_1&=F(x_1)+G^{(3)}(x_2+x_3-2x_1)\nonumber\\
	\dot{x}_2&=F(x_2)+G^{(3)}(x_1+x_3-2x_2)+G^{(3)}(x_4+x_6-2x_2)\nonumber\\
	\dot{x}_3&=F(x_3)+G^{(3)}(x_1+x_2-2x_3)+G^{(3)}(x_5+x_6-2x_3)\nonumber\\
	\dot{x}_4&=F(x_4)+G^{(3)}(x_2+x_6-2x_4)\\
	\dot{x}_5&=F(x_5)+G^{(3)}(x_3+x_6-2x_5)\nonumber\\
	\dot{x}_6&=F(x_6)+G^{(3)}(x_2+x_4-2x_6)+G^{(3)}(x_3+x_5-2x_6),\nonumber
\end{align}
and the coupling structure is captured by
\begin{align}\label{eq: incidence supp}
	I^{(3)}=\begin{pmatrix}
		1& & \\
		1&1& \\
		1& &1\\
		 &1& \\
		 & &1\\
		 &1&1  
	\end{pmatrix}.
\end{align}

To rewrite \cref{eq: node formalism} into the language of the coupled cell formalism, we use \cref{eq: incidence supp} and the homogeneity of node and edge dynamics to determine which functions should be equal (in this case, equal evolution functions for nodes $i$ and $j$ mean $\sum\limits_{k} I^{(3)}_{ik}=\sum\limits_{k} I^{(3)}_{jk}$). 
Under the formalism of Ref.\cite{golubitsky2016rigid}, the dynamics on the hypergraph shown in \cref{fig: admissible}(a) can be represented as:
\begin{align}\label{eq: GS formalism}
	\dot{x}_1&=f_1(x_1,x_2,x_3)\nonumber\\
	\dot{x}_2&=f_2(x_2,x_1,x_3,x_4,x_6)\nonumber\\
	\dot{x}_3&=f_2(x_3,x_1,x_2,x_5,x_6)\\
	\dot{x}_4&=f_1(x_4,x_2,x_6)\nonumber\\
	\dot{x}_5&=f_1(x_5,x_3,x_6)\nonumber\\
	\dot{x}_6&=f_2(x_6,x_2,x_4,x_3,x_5),\nonumber
\end{align}
In this notation, the additive structure of higher-order interactions on hypergraphs is not evident from the equations anymore.


Properties of the dyadic and higher order coupling between the nodes can be manifested in the permutational symmetries of the functions $f_i$. As mentioned, the formalism of Ref.\cite{golubitsky2016rigid} uses arrows to encode information.
Specifically, Ref.\cite{golubitsky2016rigid} states that the arrows: ``encode symmetries of those
functions, induced by permuting input arrows of the same
type.''. These symmetries of the input functions are represented with bars in the dynamical equations. However, this formalism can be ambiguous. As explained in Ref.\cite{stewart2003symmetry}, which discusses how the coupled cell network structure leads to synchronization, ``to do this, we have to order the variables suitably, and in some cases this cannot be done consistently''. 

For instance, consider the dynamics 
for the hypergraph shown in \cref{fig: admissible}(a) using the coupled cell formulation \cref{eq: GS formalism}.
The dynamics for node 1 is unambiguous. Under the undirected edge assumption, the function $f_1$ is symmetric under the permutation of $x_2$ and $x_3$, therefore the first line in \cref{eq: GS formalism} is $\dot{x}_1=f_1(x_1,\overline{x_2,x_3})$. But ambiguity arises in the dynamical equation for node 2 (as well as nodes 3 and 6).
Specifically, $f_2$ is symmetric with respect to the permutation of $x_1$ and $x_3$, as well as the permutation of $x_4$ and $x_6$, thus, in this formalism, $\dot{x}_2=f_2(x_2,\overline{x_1,x_3},\overline{x_4,x_6})$. In addition, it is symmetric to \textit{simultaneously} permuting $x_1$ and $x_4$, and $x_3$ and $x_6$. This symmetry can not be expressed with the bar formalism. More importantly, this leaves ambiguity about which arrows should be of the same type. Using the notation $a_{ij}$ for an arrow from $i$ to $j$ for convenience, it is clear that $a_{12}$ and $a_{42}$ should be of the same type, as well as $a_{32}$ and $a_{62}$, but it is not clear if these two types should be the same or distinct. This type of ambiguity will arise any time there is more than one non-dyadic hyperedge contributing to the state of a specific node. In contrast, our formalism, \cref{eq: node formalism}, while limited to specific types of dynamics, implicitly takes these properties of the dynamics into account.

Note that the issues discussed above do not arise for networks with purely dyadic interactions \cite{aguiar2018synchronization}. However, these issues can still arise for higher order coupling more general than the ones discussed in this Letter. E.g., consider the dynamics of node $i$ with contributions from triadic edges, $x_i = F(x_i)+G^{(3)}(x_i,x_j,x_k)+G^{(3)}(x_i,x_l,x_m)$, where $G^{(3)}(x_i,x_j,x_k)\neq G^{(3)}(x_i,x_k,x_j)$. Then, the dynamics of $i$ can not be written down as simply $\dot{x}_i=f(x_i,\overline{x_j,x_l},\overline{x_k,x_m})$, since the two permutations need to happen simultaneously in order to keep $f$ invariant. 

To state it more generally, the representation of higher order interactions via arrows as defined by the coupled cell formalism is appropriate when all the symmetries of functions $f$ describing the evolution of nodes $\dot{x}_i=f(x_i,x_{j_1},...,x_{j_n})$ with respect to permutations of variables $x_{j_1},...,x_{j_n}$ are defined by a symmetric group $S_{n_1}\times ... \times S_{n_k}$. Here, the actions of elements of $S_{n_l}$ are all the permutations of a subset of nodes $j_1,...,j_n$ of size $n_l$.
 
In summary, it is possible to use 
the coupled cell formalism, (e.g., \cref{eq: GS formalism}), to analyze dynamics on hypergraphs (which requires a step of obtaining whether $f_i$ and $f_j$ are the same function based on the structure of the hypergraph) together with the information about the symmetries of the functions $f_i$ to determine which states are admissible and then determine the structure of the Jacobian based on the structure of functions $f_i$. However, the bar notation that conveniently captures the permutational symmetries of the functions can fail in the case of dynamics on hypergraphs made of additive interactions of all orders. This leads to ambiguities in the visual representations via nodes and arrows in that formalism as well.
Our approach is specifically developed for dynamics in \cref{eq: dynamics} (or, more generally, \cref{eq: dynamics 2}), and therefore it implicitly takes the additivity of higher-order interactions into account. Moreover, it can be generalized to directed hypergraphs.
In addition, we note that our approach builds on recent works on simultaneous matrix block diagonalization \cite{irving2012synchronization, zhang2020symmetry}, allowing simplifications of stability analysis beyond symmetries \cite{golubitsky2016rigid}.

Very recently, the coupled cell network formalism has been applied to coupled cell hypernetworks whose coupling structure is determined by an underlying hypergraph \cite{aguiar2022network}. Using this approach the synchrony subspaces of the hypernetwork are related to balanced colorings in a corresponding incidence digraph, offering new directions for applying the coupled cell formalism to dynamics on hypergraphs.

\end{document}